\documentclass[conference]{IEEEtran}
\IEEEoverridecommandlockouts
\usepackage{cite}
\usepackage{amsmath,amssymb,amsfonts}
\usepackage{algorithmic}
\usepackage{graphicx}
\usepackage{textcomp}
\usepackage{xcolor}
\usepackage{tabularx}
\usepackage[hyphens,spaces,obeyspaces]{url}
\def\BibTeX{{\rm B\kern-.05em{\sc i\kern-.025em b}\kern-.08em
    T\kern-.1667em\lower.7ex\hbox{E}\kern-.125emX}}

\begin{document}

\title{Web Links Prediction And Category-Wise Recommendation Based On Browser History}
\author{
    \IEEEauthorblockN{Ashadullah Shawon,  Syed Tauhid Zuhori , Firoz Mahmud , Md. Jamil-Ur Rahman}
    \IEEEauthorblockA{ Rajshahi University of Engineering and Technology
    \\\{shawonashadullah,tauhid.ruet04,fmahmud.ruet,jamilruet13\}@gmail.com}
}

\maketitle

\begin{abstract}
A web browser should not be only for browsing web pages but also help users to find out their target websites and recommend similar type websites based on their behavior. Throughout this paper, we propose two methods to make a web browser more intelligent about link prediction which works during typing on address-bar and recommendation of websites according to several categories. Our proposed link prediction system is actually frecency prediction which is predicted based on the first visit, last visit and URL counts. But recommend system is the most challenging as it is needed to classify web URLs according to names without visiting web pages. So we use existing model for URL classification. The only existing approach gives unsatisfactory results and low accuracy. So we add hyperparameter optimization with an existing approach that finds the best parameters for existing URL classification model and gives better accuracy. In this paper, we propose a category wise recommendation system using frecency value and the total visit of individual URL category.
\end{abstract}


\section{Introduction}
Nowadays most important features of different kinds of websites are quick searching, content recommendation and tracking user behavior. The advantages of these features are easily finding target contents of a user. Web contents are recommended based on content tags. These tags define the category and identity of those contents. For quick searching, websites analyze search keywords and predict target search content. Web-browser also deals with many websites. For quick searching or link prediction on address-bar, web-browser analyze the history of a user. Web-browser history contains much information but most important information are URL, first-visit, last-visit, click-counts and frecency. Learning an implicit algorithm can make the browser more intelligent. So supervised learning algorithm is used for link prediction in this paper. Regression method is used for predicting frecency value. Thus higher frecency value means the higher interest of a user~\cite{ref_url}. But recommendation system for a browser is difficult than web contents as for a recommendation, the browser needs to classify the URLs. Classifying the URLs based on their name is a challenging task.

\section{Related Work}
There are some research papers which were published earlier based on either browser history or web URL classification. An earlier research showed the method and system for auto-generating web URL list during searching based on browser history~\cite{ref_patent1}. The existing system of link prediction of a web-browser follows explicit algorithm~\cite{ref_url1}. For URL classification most of the previous paper focused on feature extraction techniques, some paper focused on models~\cite{ref_article1}~\cite{ref_article2}. Macro-averaged F-measure of URL classification is obtained in another research using Naive Bayes and SVM algorithm~\cite{ref_article22}. In 2018, we published a research paper~\cite{ref_article222} focusing on only website or URL classification. This research paper achieved the best accuracy so far. We used balanced DMOZ dataset there and took 2000 samples per class. But in this research paper, we are using unbalanced and large samples of DMOZ dataset.

\section{Proposed Method}
In our research, we approach for both link prediction and web URL classification. We modify current link prediction system and add some techniques for better test accuracy for URL classification. Finally, we use the result of both link prediction and URL classification to create a ranking of websites and recommendation. Fig. 1 shows the diagram of our proposed method. We use browser history dataset to predict frecency value. As it is a supervised learning method, at first machine is trained by train dataset to predict frecency using linear regression algorithm.
\begin{figure}[htbp]
\includegraphics[width=1.00\columnwidth]{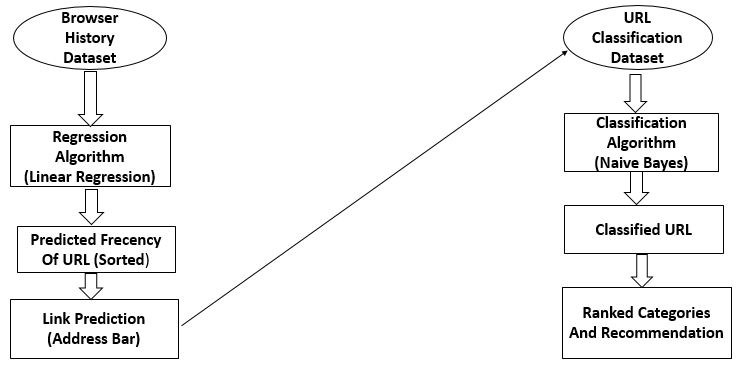}
\caption{Proposed Method.} \label{fig1}
\end{figure}
After predicted frecency value, it is known that which URLs have higher frecency that means higher user interest. But for a recommendation, it is needed to identify the category of those URL so that browser can fetch similar category type links or news. So another important part of our proposed method is classifying URL. A machine is trained from URL classification dataset using multinomial Naive Bayes classification algorithm~\cite{ref_article22}. Then our system classifies URL and finds the most visited category of websites. Finally, our system recommends websites users according to the ranking of individual category.

\section{Frecency Prediction}
\subsection{Browser History Dataset}
Browser history dataset is a real-life browsing data that is stored by the browser. We have collected it from Mozilla Firefox web browser and it contains much information such as URL, first visit time, last visit time, URL counts, URL lengths, frecency, URL title etc. There are total 2305 URL. We need URL, first visit time, last visit time, URL counts and frecency as our interest is related to time and URL counts. Time is converted to Unix timestamp for calculation. Thus after preprocessing, sample dataset looks like Table I:
\begin{table}[htbp]
\begin{center}
\caption{Sample Browser history dataset}\label{tab1}
\resizebox{\columnwidth}{!}{\begin{tabular}{|l|l|l|l|l|}
\hline
URL &  First Time Visit & Last Time Visit & URL Counts & Frecency\\
\hline
https://web.facebook.com/ & 1521241972 & 1522351859 & 177 & 56640\\
http://localhost/phpmyadmin/ & 1518413861 & 1522075694 & 24 & 39312\\
https://mail.google.com/mail/u/ & 1516596003 & 1522352010 & 36 & 33264\\
https://github.com/ & 1517215489 & 1522352266 & 37 & 27528\\
https://www.youtube.com/ & 1517229227 & 1521978502 & 24 & 14792\\
\hline
\end{tabular}}
\end{center}
\end{table}

\subsection{Current Frecency Calculation System}
Current Frecency System is not being calculated by a learning algorithm. According to Mozilla developer blog: ``Frecency is a score given to each unique URI in Places, encompassing bookmarks, history and tags. This score is determined by the amount of re-visitation, the type of those visits, how recent they were, and whether the URI was bookmarked or tagged. The word frecency itself is a combination of the words frequency and recency. The default frecency value for all valid entries is -1. Places with this value can show up in autocomplete results. Invalid places have a frecency value of zero, and will not show up in autocomplete results. Examples of invalid places are place: queries, and unvisited livemark feed entries"~\cite{ref_url}. Frecency rating is being calculated currently by the exponential decay. Decay rate constant $\lambda$, points of visit $p$, age of visit $a$ that is difference between last and first visit are needed to calculate current value of a visit $\beta$ and frecency score $\gamma$. So Eq.1, Eq.2, Eq.3 show the current calculation system~\cite{ref_url1}.
\begin{center}
\begin{equation}\label{lm}
\lambda=\frac{ln(2)}{30}
\end{equation}
\end{center}
\begin{center}
\begin{equation}\label{cv}
\beta=p*e^{-\lambda*a}
\end{equation}
\end{center}
\begin{center}
\begin{equation}\label{fs}
\gamma=\sum_{i=1}^{n}\beta
\end{equation}
\end{center}
\subsection{Our Proposed Frecency Prediction System}
We analyze browser history and find some features that are very important for frecency prediction. From browser history dataset our columns are URL, first-visit, last-visit, click-counts and frecency. We want to train browser from a history dataset so that it can predict frecency for new history dataset. So we propose to use a supervised learning algorithm to predict frecency value of unknown URLs implicitly instead of calculating by an explicit algorithm. As a result, browser need not calculate frecency explicitly. According to browser history, we observe the relation between frecency value and URL, first-visit, last-visit, click-counts for frecency value prediction.
So features of the dataset are first-visit, last-visit and click-counts. URL is not necessary during prediction, because frecency is changed according to first-visit, last-visit and click-counts. But URL is final output for link prediction. So URL column should be present. Our target column is frecency.\newline\newline
\textbf{Features X :}
\begin{itemize}
  \item First Visit Time
  \item Last Visit Time
  \item Click Counts
\end{itemize}
\textbf{Target Y :}
\begin{itemize}
  \item Frecency
\end{itemize}
After observing the relationship in Fig.2, we have found a linear line. Linear line is fitted with some points.\newline
 \begin{figure}[htbp]
  \centering
  \includegraphics[width=1.00\columnwidth]{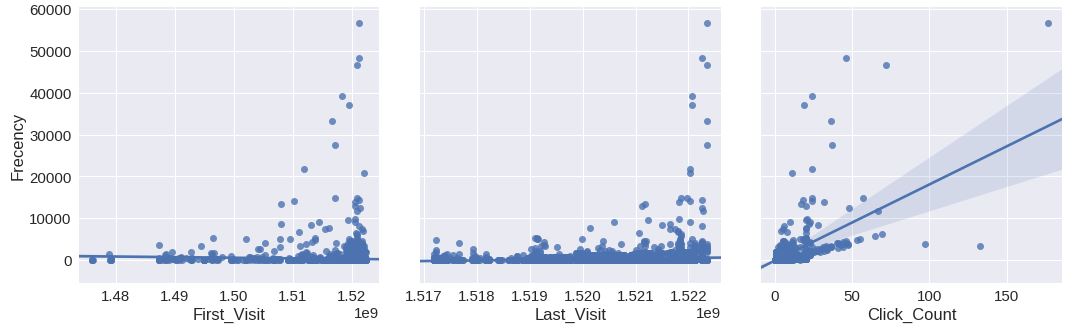}
  \caption{Relationship Between Features And Target}\label{fig2}
\end{figure}
Finally, after observing linear relationship, we propose to predict frecency by the multivariate linear regression algorithm.
\subsection{Multivariate Linear Regression Algorithm:}
Linear regression with multiple variables is called multivariate linear regression. A common method for multiple linear regression~\cite{ref_article23}~\cite{ref_book} is shown in Eq.4. Eq.5 finds the minimized cost function $J(\theta)$ and Eq.7 finds all values of the $\theta$ vector~\cite{ref_url2}.
\begin{equation}\label{Hypothesis}
h=\theta_{0}+\theta_{1}x_{1}+\theta_{2}x_{2}+\theta_{3}x_{3}+.....+ \theta_{m}x_{m}=\theta^{T}X
\end{equation}
Here, $h$ is hypothesis. So $\theta_{0},\theta_{1},\theta_{2}..........\theta_{m}$ is calculated such that cost function $J(\theta_{0},\theta_{1},\theta_{2}.......\theta_{m})$ is minimized. So cost function equation is:
\begin{equation}\label{cost}
J(\theta)=\frac{1}{2m}\sum_{i=1}^{m}(h_{\theta}(x^{i})-y^{i})^{2}
\end{equation}
$\theta$ is a column vector here.
\begin{center}
\begin{equation}\label{tm}
\theta=\left(
         \begin{array}{c}
           \theta_{0} \\
           \theta_{1} \\
           .\\
           .\\
           \theta_{m} \\
         \end{array}
       \right)
\end{equation}
\begin{equation}\label{theta}
\theta=(X^{T}X)^{-1}X^{T}Y
\end{equation}
\end{center}
After applying $\theta$ values in Eq.(4), predicted value can be achieved as $h$ is our hypothesis.
\section{URL Classification}
\subsection{Dataset Visualization}
We use Open Directory Project DMOZ dataset for URL classification~\cite{ref_url3}. The dataset is divided into training set and test set. We also include some extra entries in DMOZ test dataset. For 4 class URL classification, the training dataset visualization is given below:
 \begin{figure}[htbp]
  \centering
  \includegraphics[width=1.00\columnwidth]{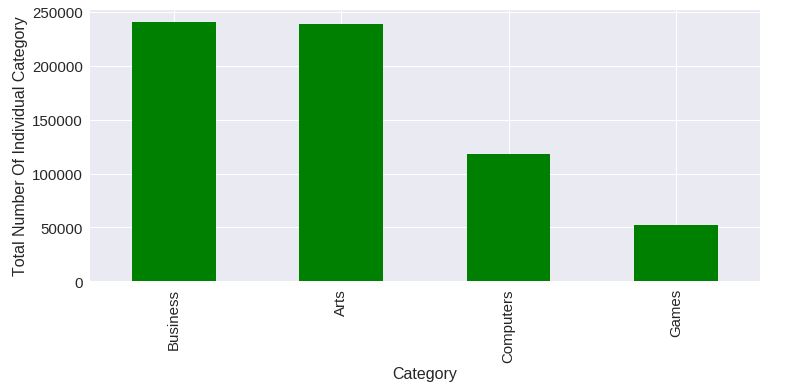}
  \caption{DMOZ Dataset Visualization For 4 Classes}\label{fig3}
\end{figure}
There are total 650000 entries and 15 categories. The dataset split is about 70\% training  and 30\% testing with 10 fold cross validation. The categories are Adults, Arts, Business, Computers, Games, Health, Home, News, Recreation, Reference, Science, Society, Shopping, Sports, and Regional. We Table II shows a small part of DMOZ dataset.
\begin{table}[htbp]
\caption{DMOZ dataset sample.}\label{tab2}
\begin{center}
\begin{tabular}{|l|l|}
\hline
URL &  Category\\
\hline
http://www.gamespot.com/ps/ & Games\\
https://www.gamefun.com 	& Games\\
https://drive.google.com/drive/my-drive & Computers\\
http://www.narutofan.com/ & Arts\\
\hline
\end{tabular}
\end{center}
\end{table}

\subsection{Current Vs Proposed URL Classification System}
The first difference between previous approach [4] and our approach is testing data. The previous researchers took only 15000 testing data for verifying testing accuracy but our testing data is about 195000 which is a massive testing dataset. Current URL classification system is based on n-gram language model and multinomial Naive Bayes classifier~\cite{ref_article1}. Though n-gram language model and multinomial Naive Bayes classifier are the main models of URL classification, it is needed to add some technique to increase the performance of existing model. We use an existing approach in our research and we add hyperparameter optimization for better accuracy. There are many hyperparameter tuning methods. We select random search hyperparameter optimization method. Because random search can find better parameters for a model within a small fraction of computation time and make different parameters important on different data sets~\cite{ref_article3}. It is performed by evaluating n uniformly random points in the hyperparameter space and select the one producing the best performance~\cite{ref_url4}. According to another paper of random search algorithm, the update procedure to find best parameters is shown in Eq.8 where $X_{k+1}$ is current point and $V_{k+1}$ is candidate point~\cite{ref_article4}.\newline
\begin{center}
\begin{equation}
  X_{k+1} =
  \begin{cases}
    V_{k+1} & \text{if $f(V_{k+1})<f(X_{k})$} \\
    X_{k} & \text{otherwise}
  \end{cases}
\end{equation}
\end{center}
If this procedure trap in a local optimum then the solution is simulated annealing~\cite{ref_article5}. The solution procedure is shown in Eq.9.
\begin{center}
\begin{equation}
  X_{k+1} =
  \begin{cases}
    V_{k+1} & \text{with probability min$\Big\{1,e(\frac{f(X_{k})-f(V_{k+1})}{T_{k}})\Big\}$} \\
    X_{k} & \text{otherwise}
  \end{cases}
\end{equation}
\end{center}
\section{Experimental Results}
\subsection{Frecency Prediction}
Our target is to sort the links according to users interest so that users will be suggested to click those links that are in top position. Experimental results of frecency prediction are shown in Table III. During calculating results, we use another external history dataset to avoid bias problem. We measure mean square error, root mean square(RMS) error and score which is the coefficient of determination $R^2$ of the prediction.
\begin{table}[htbp]
\setlength{\tabcolsep}{0.9em}
\renewcommand{\arraystretch}{1.9}
\caption{Experimental results of frecency prediction.}\label{tab3}
\begin{center}
\begin{tabular}{|l|l|l|}
\hline
Mean Square Error &  Root Mean Square Error & Score\\
\hline
586430.6574 & 765.7876 & 87.62\%\\
\hline
\end{tabular}
\end{center}
\end{table}\newline
Predicted relationship vs actual relationship is shown in Figure 4. We observe that predicted relationship line is linear but actual relationship line is slightly bent. So according to Figure 4, the predicted result is not individually accurate but  relatively maximum predicted frecency values are sorted as actual frecency values. Thus this result is meaningful for detecting most interesting links.
\begin{figure}[htbp]
  \centering
  \includegraphics[scale=0.5]{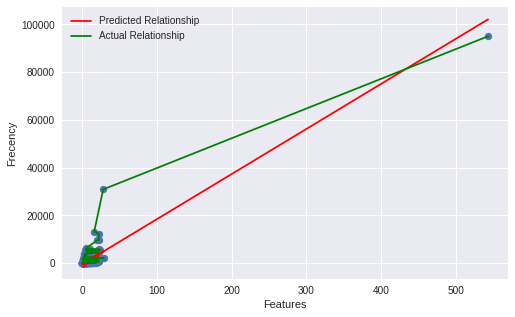}
  \caption{Predicted relationship vs Actual relationship}\label{fig5}
\end{figure}
\subsection{Link Prediction}
After getting predicted frecency value, we sort the URL list according to higher frecency. Table IV shows the sample sorted result with predicted frecency. The link prediction for the search term ``loc" will predict Table IV URLs. We develop a live link prediction system for this result only~\cite{ref_url5}.
\begin{table}[htbp]
\begin{center}
\caption{Predicted Frecency.}\label{tab4}
\begin{tabular}{|l|l|l|}
\hline
URL &  URL Counts & Frecency \\
\hline
http://localhost/phpmyadmin/ & 16 & 2906.7627\\
http://localhost:8888/tree & 15 & 2717.497\\
http://localhost:8000/home & 13 & 2274.1109\\
\hline
\end{tabular}
\end{center}
\end{table}
\subsection{URL Classification}
Previous approach[4] did not find best parameters for URL classification. We find the best parameters for existing n-gram model and improve the accuracy of previous research. We get the best parameters and best mean value is 0.69971. The best parameters are shown in Table V and all parameters list  is shown in Table VI. The best parameters are found by random search analyzing 8 iterations. Here clf-alpha is a learning rate.
\begin{table}[htbp]
\setlength{\tabcolsep}{0.4em}
\renewcommand{\arraystretch}{1.1}
\caption{Best Parameters}\label{tab5}
\begin{center}
\begin{tabular}{|l|l|l|l|l|}
\hline
mean & standard-deviation & n-gram-range &  tfidf-use-idf & clf-alpha\\
\hline
0.69971 &  0.00035 & (1,2) & True & 0.01\\
\hline
\end{tabular}
\end{center}
\end{table}
\begin{table}[htbp]
\setlength{\tabcolsep}{0.5em}
\renewcommand{\arraystretch}{1.5}
\caption{All Parameters list by random search}\label{tab6}
\begin{center}
\begin{tabular}{|l|l|l|l|l|}
\hline
mean & standard-deviation & n-gram-range &  tfidf-use-idf & clf-alpha\\
\hline
0.69245 &  0.00036 & (1,1) & True & 0.01\\
0.69971 &  0.00035 & (1,2) & True & 0.01\\
0.69460 &  0.00053 & (1,1) & False & 0.01\\
0.69702 &  0.00047 & (1,2) & False & 0.01\\
0.69153 &  0.00028 & (1,1) & True & 0.001\\
0.69804 &  0.00034 & (1,2) & True & 0.001\\
0.69348 &  0.00062 & (1,1) & True & 0.001\\
0.69614 &  0.00047 & (1,2) & True & 0.001\\
\hline
\end{tabular}
\end{center}
\end{table}
Table VII shows the Precision, Recall, F1 score and Accuracy for 2, 4 and 15 class URL classification system without applying random search of previous approach. Table VIII shows the better result than Table VII after applying random search of our proposed method.
\begin{table}[!htb]
\setlength{\tabcolsep}{0.5em}
\renewcommand{\arraystretch}{1.5}
\caption{Classification Accuracy Without Random Search (Previous Method)}\label{tab7}
\begin{center}
\begin{tabular}{|l|l|l|l|l|}
\hline
Class & Precision & Recall & F1 score & Accuracy \\
\hline
2 & 84.35\% & 83.33\% & 83.22\% & 83.33\%\\
4 & 73.45\% & 68.33\% & 66.29\% & 68.33\%\\
15 & 61.66\% & 44.01\% & 39.74\% & 44.01\%\\
\hline
\end{tabular}
\end{center}
\end{table}
\begin{table}[!htb]
\setlength{\tabcolsep}{0.5em}
\renewcommand{\arraystretch}{1.9}
\caption{Classification Accuracy With Random Search (Proposed Method)}\label{tab8}
\begin{center}
\begin{tabular}{|l|l|l|l|l|}
\hline
Class & Precision & Recall & F1 score & Accuracy \\
\hline
2 & 85.70\% & 84.34\% & 84.20\% & 84.34\%\\
4 & 73.74\% & 71.23\% & 70.54\% & 71.23\%\\
15 & 56.34\% & 49.87\% & 48.66\% & 49.87\%\\
\hline
\end{tabular}
\end{center}
\end{table}
Comparing Table VII and Table VIII, it is clear that our approach gives a better result than previous.
\subsection{Recommendation}
Recommendation result is highly dependent on URL classification. Because recommendation URLs are fetched according to URL classification output. After applying both frecency prediction and URL classification we get the predicted frecency value and predicted URL category for browser history test dataset. Some of the predicted sample results for test set is shown in Table IX. Table IX result is sorted by higher frecency value so that during typing on the address bar of a browser, a user will be suggested the higher frecency URL based on user typed string. Finally, we get the total number of visit according to individual categories from browser history and Figure 5 shows the total visit number according to categories.
\begin{table}[htbp]
\begin{center}
\caption{Predicted sample result for frecency and category}\label{tab9}
\resizebox{\columnwidth}{!}{\begin{tabular}{|l|l|l|l|}
\hline
URL &   URL Counts & Frecency & Category\\
\hline
https://web.facebook.com/  & 543 & 102108.26 & Computers\\
https://drive.google.com/drive/my-drive & 28 & 4873.655 & Computers\\
http://codeforces.com/contests & 21 & 3650.896 & Arts\\
https://www.floydhub.com/jobs & 4 & 665.371 & Business\\
http://www.cricbuzz.com/live-cricket-scores & 4 & 579.825 & Games\\
http://localhost/map/googlemap.php & 9 & 528.395 & Computers\\
https://www.kaggle.com/competitions & 2 & 309.769 & Arts\\
https://freebitco.in/ & 1 & 111.909 & Business\\
\hline
\end{tabular}}
\end{center}
\end{table}\break
\begin{figure}[htbp]
  \centering
  \includegraphics[width=1.00\columnwidth]{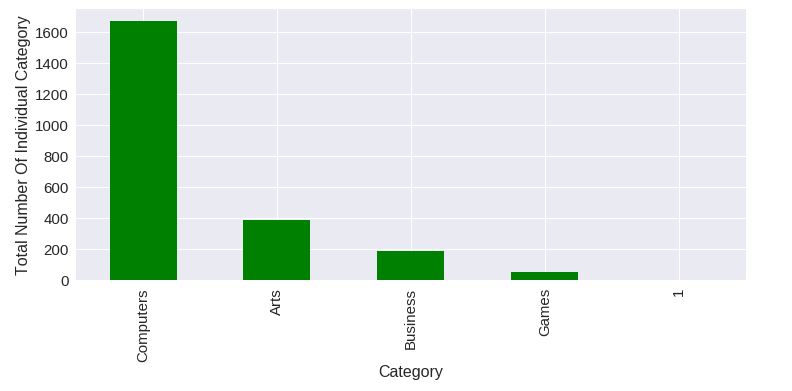}
  \caption{Total visit number according to categories}\label{fig5}
\end{figure}
We find the total frecency T of every common category from Table IX and the total visit of individual category.
\begin{center}
\begin{equation}\label{rn}
T=\sum_{i=1}^{k} f_{i}
\end{equation}
\end{center}
So the probability of individual category is defined in Eq. 11.
\begin{center}
\begin{equation}\label{pr}
P(T_{i})=\frac{T_{i}}{\sum_{i=1}^{t}T_{i}}
\end{equation}
\end{center}
Now, according to the probability of individual category, we get the ranking of categories. Table X shows the ranking of categories.
\begin{table}[htbp]
\setlength{\tabcolsep}{0.9em}
\renewcommand{\arraystretch}{2.1}
\caption{Ranking of categories}\label{tab10}
\begin{center}
\begin{tabular}{|l|l|l|}
\hline
Rank & Category & Probability\\
\hline
1 & Computers & 0.77 \\
\hline
2 & Arts & 0.097 \\
\hline
3 & Business & 0.074 \\
\hline
4 & Games & 0.051 \\
\hline
\end{tabular}
\end{center}
\end{table}
So we find the most interesting category of this browser history is ``Computers". According to ranking, the similar type of websites are fetched from the database or scraped from the internet in real time. Table XI and XII shows the sample recommendation based on user interest. We develop a live sample recommendation result for better understanding~\cite{ref_url6}.
\begin{table}[htbp]
\setlength{\tabcolsep}{0.5em}
\renewcommand{\arraystretch}{1.5}
\caption{Sample recommendation for most interested category}\label{tab11}
\begin{center}
\begin{tabular}{|l|l|}
\hline
Computers Category\\
\hline
https://twitter.com  \\
\hline
https://bitbucket.org\\
\hline
https://reddit.com \\
\hline
https://instagram.com \\
\hline
https://datascience.com \\
\hline
https://khanacademy.org \\
\hline
https://www.computer.org\\
\hline
https://www.apple.com\\
\hline
https://www.ieee.org\\
\hline

\end{tabular}
\end{center}
\end{table}
\begin{table}[htbp]
\setlength{\tabcolsep}{0.5em}
\renewcommand{\arraystretch}{1.9}
\caption{Sample recommendation for others category}\label{tab12}
\begin{center}
\resizebox{\columnwidth}{!}{\begin{tabular}{|l|l|l|}
\hline
Arts Category & Business Category & Games Category\\
\hline
https://cartoonnetwork.com & https://www.business.gov.au/ & http://www.mariogames.be\\
\hline
https://watchcartoononline.io & https://www.linkedin.com/jobs & https://www.goal.com/en\\
\hline
http://ben10.wikia.com/wiki/ & https://www.upwork.com & http://gamesgames.com\\
\hline
\end{tabular}}
\end{center}
\end{table}
According to all results, we find an intelligent link prediction and a recommendation system  for a web browser.
\section{Conclusion}
In this paper, we have focused on implicit prediction system for frecency calculation instead of the explicit algorithm and random search algorithm for improving the accuracy of web classification as well as the recommendation system. We have shown an intelligent link prediction method and improvement of URL classification accuracy for several classes. But as a future work, frecency prediction should be more accurate and URL classification accuracy need to be more improved. We believe that polynomial regression or more advanced regression can reduce root mean square error and improve frecency prediction. More advanced parameter optimization and adding web page title feature in DMOZ dataset may increase accuracy for multi-class URL classification. Finally, we recommend to continue our research in that direction.

\end{document}